\documentclass[]{bytedance_seed}

\usepackage[toc,page,header]{appendix}

\usepackage{minitoc}
\usepackage{caption}
\usepackage{cuted}
\usepackage{cleveref}
\usepackage{diagbox}
\usepackage{paralist}
\usepackage{enumitem}
\usepackage{listings}
\usepackage{listings-rust}
\definecolor{listing-background}{HTML}{fcfcfc}
\definecolor{listing-rule}{HTML}{c9c9c9}
\definecolor{listing-numbers}{HTML}{575757}
\definecolor{listing-text-color}{HTML}{3b3b3b}
\definecolor{listing-keyword}{HTML}{1c1c1c}
\definecolor{listing-keyword-2}{HTML}{1e1e1e} 
\definecolor{listing-keyword-3}{HTML}{1e1e1e} 
\definecolor{listing-identifier}{HTML}{2e2e2e}
\definecolor{listing-string}{HTML}{171717}
\definecolor{listing-comment}{HTML}{757575}

\lstdefinestyle{custom_listing_style}{
  numbers          = left,
  xleftmargin      = 2.7em,
  framexleftmargin = 2.5em,
  basicstyle       = \color{listing-text-color}\linespread{1.0}%
                      \lst@ifdisplaystyle%
                      \scriptsize%
                      \fi\ttfamily{},
  captionpos       = b,
  breaklines       = true,
  frame            = none,
  framesep         = 0.19em,
  rulecolor        = \color{listing-rule},
  frameround       = ffff,
  tabsize          = 4,
  numberstyle      = \color{listing-numbers},
  aboveskip        = 0.1em,
  belowskip        = 0.1em,
  abovecaptionskip = 0em,
  belowcaptionskip = 0.1em,
  keywordstyle     = {\color{listing-keyword}\bfseries},
  keywordstyle     = {[2]\color{listing-keyword-2}\bfseries},
  keywordstyle     = {[3]\color{listing-keyword-3}\bfseries\itshape},
  sensitive        = true,
  identifierstyle  = \color{listing-identifier},
  commentstyle     = \color{listing-comment},
  stringstyle      = \color{listing-string},
  showstringspaces = false,
  escapeinside     = {/*@}{@*/}, 
  literate         =
  {á}{{\'a}}1 {é}{{\'e}}1 {í}{{\'i}}1 {ó}{{\'o}}1 {ú}{{\'u}}1
  {Á}{{\'A}}1 {É}{{\'E}}1 {Í}{{\'I}}1 {Ó}{{\'O}}1 {Ú}{{\'U}}1
  {à}{{\`a}}1 {è}{{\`e}}1 {ì}{{\`i}}1 {ò}{{\`o}}1 {ù}{{\`u}}1
  {À}{{\`A}}1 {È}{{\`E}}1 {Ì}{{\`I}}1 {Ò}{{\`O}}1 {Ù}{{\`U}}1
  {ä}{{\"a}}1 {ë}{{\"e}}1 {ï}{{\"i}}1 {ö}{{\"o}}1 {ü}{{\"u}}1
  {Ä}{{\"A}}1 {Ë}{{\"E}}1 {Ï}{{\"I}}1 {Ö}{{\"O}}1 {Ü}{{\"U}}1
  {â}{{\^a}}1 {ê}{{\^e}}1 {î}{{\^i}}1 {ô}{{\^o}}1 {û}{{\^u}}1
  {Â}{{\^A}}1 {Ê}{{\^E}}1 {Î}{{\^I}}1 {Ô}{{\^O}}1 {Û}{{\^U}}1
  {œ}{{\oe}}1 {Œ}{{\OE}}1 {æ}{{\ae}}1 {Æ}{{\AE}}1 {ß}{{\ss}}1
  {ç}{{\c c}}1 {Ç}{{\c C}}1 {ø}{{\o}}1 {å}{{\r a}}1 {Å}{{\r A}}1
  {€}{{\EUR}}1 {£}{{\pounds}}1 {«}{{\guillemotleft}}1
  {»}{{\guillemotright}}1 {ñ}{{\~n}}1 {Ñ}{{\~N}}1 {¿}{{?`}}1
  {…}{{\ldots}}1 {≥}{{>=}}1 {≤}{{<=}}1 {„}{{\glqq}}1 {“}{{\grqq}}1
  {”}{{''}}1
}
\usepackage{makecell}
\usepackage{mathtools}
\usepackage{multicol}
\usepackage{multirow}
\usepackage{url}
\usepackage{subcaption}
\usepackage{tablefootnote}
\usepackage{xcolor}
\usepackage{xspace}
\usepackage{mathtools}

\newcommand{\cleanmap}[0]{\xmapsto{clean}}

\renewcommand{\paragraph}[1]{\vspace{0.02in}\noindent{\bf #1}}

\crefformat{section}{\S#2#1#3} 
\crefformat{subsection}{\S#2#1#3}
\crefformat{subsubsection}{\S#2#1#3}

\newcommand{\name}[0]{\textsc{GraphGuard}\xspace}
\newcommand{\egg}[0]{\texttt{egg}\xspace}

\newcommand{\enodes}[0]{\texttt{ENode}s\xspace}
\newcommand{\eclass}[0]{\texttt{EClass}\xspace}
\newcommand{\eclasses}[0]{\texttt{EClass}es\xspace}
\newcommand{\egraph}[0]{\texttt{EGraph}\xspace}
\newcommand{\egraphs}[0]{\texttt{EGraphs}\xspace}
\newcommand{\company}[0]{Bytedance\xspace}

\newcommand{\eat}[1]{}
\newcommand{\allnotes}[1]{\textit{#1}}

\newcommand{\notezhanghan}[1]{\allnotes{{\it\color{blue}[ZH: #1]}}}

\newcommand{\noteding}[1]{\allnotes{{\it\color{purple}[DD: #1]}}}

\newcommand{\func}[2]{\text{#1}#2}
\definecolor{coral}{RGB}{252, 236, 234}
\definecolor{lgreen}{RGB}{224, 250, 227}

\newcommand{\figuresize}[0]{\linewidth/2}

\title{ Verify Distributed Deep Learning Model Implementation Refinement with Iterative Relation Inference }

\author[2,*,\ddagger]{Zhanghan Wang}
\author[2,\ddagger]{Ding Ding}
\author[1,\dagger]{Hang Zhu}
\author[1,\dagger]{Haibin Lin}
\author[2,\dagger]{Aurojit Panda}

\affiliation[1]{ByteDance Seed}
\affiliation[2]{New York University}

\contribution[*]{Work done at ByteDance Seed}
\contribution[\dagger]{Corresponding authors}
\contribution[\ddagger]{Equal contribution}

\abstract{
Distributed machine learning training and inference is common today because today's large models require more memory and compute than can be provided by a single GPU. Distributed models are generally produced by programmers who take a sequential model specification and apply several distribution strategies to distribute state and computation across GPUs. Unfortunately, bugs can be introduced in the process, and a distributed model implementation's outputs might differ from the sequential model's outputs. In this paper, we describe an approach to statically identify such bugs by checking \emph{model refinement}, that is, can the sequential model's outputs be reconstructed from the distributed model's outputs? Our approach, implemented in \name, uses iterative rewriting to prove model refinement. Our approach can scale to today's large models and deployments: we evaluate it using GPT and Llama-3.
Further, it provides actionable output that aids in bug localization.  
}

\date{\today}

\begin{document}
\maketitle


\section{Introduction}
Large machine learning models require more memory than is available on any single GPU. Furthermore, training them or using them for inference requires significant compute capacity, making it infeasible to use a single GPU. Consequently, it is now the norm to deploy these models on multiple GPUs, spread across multiple servers, for training and inference tasks. The approach taken when implementing a distributed ML model has a significant impact on resource efficiency and performance, and thus several distribution strategies~\cite{megatron-lm-1, megatron-lm-2, megatron-lm-3, deepspeed,alpa,nnscaler} are used when implementing ML models. However, implementing distributed models requires programmer effort, and bugs can be introduced during implementation.

To see why bugs can be introduced, we start by looking at a common workflow for creating a distributed model implementation: First, an ML model architect specifies a model architecture as a series of operations. This architecture specification is sequential, i.e., it is written assuming that the operations run on a single GPU (or processor) and operate on local data. Next, an implementer converts the specification into a distributed version by deciding how to partition model state and computation. When doing this, the implementer needs to add communication and transformation operations to preserve the sequential specification semantics. Unfortunately, an implementer might use incorrect parameters (e.g., incorrectly specifying padding or offsets, or using the wrong scaling factor, see \S\ref{sec:case-study}), when adding these additional operations, or worse forget some, resulting in bugs.

Indeed, as we show in \S\ref{sec:case-study}, at \company we found that several bugs had been introduced when creating a distributed implementation for a recent ML model architecture. Recent work~\cite{nnscaler, vescale, aerify-euromlsys25},
has similarly found bugs in open-source distributed ML model implementations. 

In this paper, our goal is to identify bugs introduced when implementing distributed ML models, before they are deployed. To do so, we propose a static approach for checking \emph{model refinement} (\S\ref{sec:design:term}): that is, can a sequential model $G_s$'s outputs be reconstructed from the outputs of a distributed model implementation $G_d$? In developing this approach, we had to address two core challenges: \emph{scalability}: ML models are growing in size and the number of GPUs used by implementations is also growing, and we aim for approaches that can be applied to today's models and implementations; and \emph{usability}: we want to provide the users of our approach with actionable information that can help address bugs.

Our approach, which we have implemented in a tool called \name (\S\ref{sec:design}), uses iterative term rewriting to generate a relation (\S\ref{sec:design:term}) to map the outputs produced by the implementation $G_d$ to $G_s$'s outputs. $G_d$ refines $G_s$ if \name can find a \emph{complete clean relation}, i.e., a relation that can be used to reconstruct all outputs from $G_s$ without requiring additional computation (beyond what is required to gather and combine outputs from multiple GPUs). The lack of a clean relation indicates a bug. \name can work with models and implementation written in PyTorch and other popular frameworks without requiring significant effort, and is thus easily applied to existing distributed ML model implementations.

We address the scalability and usability challenge by adopting an iterative approach where each operator in $G_s$ is processed individually (\S\ref{sec:step}). Processing a single operator limits the number of rewritten terms that \name has to consider, and that the runtime grows linearly with model complexity. Our evaluation \S\ref{sec:scalability} shows that this approach allows us to be used to check implementations of state-of-the-art models (e.g., GPT, Qwen2, Llama-3). It takes \name between 6---167 seconds (or less than 3 minutes) to check these models. In terms of usability, because \name processes a single operator at a time, its output aids programmes in localizing and addressing the bugs it identifies (\S\ref{sec:case-study}).

Furthermore, as we discuss in \S\ref{s:guarantees}, processing individual $G_s$ operators does not affect \name's soundness: \name will always report bugs if they exist. However, our approach is based on observations about how programmers (and compilers) translate model specifications to distributed implementations, and \name cannot ensure completeness if these observations do not hold, i.e., in some cases \name might raise a false alarm and report that a correct implementation is buggy. However, we did not encounter false alarms when using \name and evaluating it on open-source and proprietary models.

\section{Background}\label{s:background}

\subsection{Distribution Strategies}
We start by briefly reviewing the strategies used to distribute (aka parallelize) machine learning models. These strategies dictate how inputs are partitioned across GPUs (name users must provide these as input, $R_i$) and how outputs at multiple GPUs can be combined to recover the original sequential model's output (\name outputs this if it can prove model refinement).

A correct distribution strategy ensures that if inputs are partitioned correctly (the input relation holds) then combining the output using the approach provided by the strategy will recover the original result. This observation motivated our formulation of the model refinement problem: we check that there is a mapping from the distributed model's output to the sequential model's output assuming that a user-provided input mapping is correct.

\paragraph{Data Parallelism (DP)} was an early distribution strategy to improve training performance. When using this strategy, each GPU (or rank) runs the same model implementation, and independently computes a gradient. This strategy requires that training data be partitioned across GPUs (or ranks), and ensures that using gradients aggregated from these machines (using all-reduce) produces the same training result as training on a single machine.

\paragraph{Tensor Parallelism (TP)~\cite{narayanan2021efficient,korthikanti2023reducing}, Sequence Parallelism (SP)~\cite{megatron-lm-3} and Context Parallelism (CP)~\cite{nvidia-context-parallelism,dubey2024llama,liu2023ringattentionblockwisetransformers}} partition one or more operators across multiple GPUs. These strategies require that input tensors (TP), sequences (SP) or context (CP) be partitioned across GPUs, and they specify what operations should be used to combine their outputs. They ensure that assuming inputs are partitioned correctly, the combined output is the same (or produces the same results) as the operator(s) running on a single GPU.

\paragraph{Expert Parallelism (EP)}~\cite{switch-transformers, fastmoe} is a distribution strategy targeting mixture-of-experts (MoE) models. These models consist of multiple experts, and expert parallelism distributes experts across GPUs. This strategy requires that inputs be routed to the distributed experts using the same routing mechanism as would be used in a sequential implementation, and uses the same operators as the sequential implementation to combine outputs, ensuring that distributed and sequential models have the same output.

\paragraph{Pipeline Parallelism (PP)} is a parallelism approach with several variants~\cite{huang2019gpipe, fan2021dapple,deepseekai2025deepseekv3technicalreport}, all of which partition the model's layers across multiple GPUs. PP requires input batches to be partitioned into microbatches, and combines the output using gradient accumulation. Similar to TP, it ensures that the accumulated result is the same as would be expected from running the model on a single GPU.

As can be observed, all six distribution strategies provide similar correctness guarantees: if the strategy is correctly applied to a sequential model $G_s$ to produce a distributed implementation $G_d$, and if the strategy's input relation is used to map sequential inputs to $G_d$'s inputs, then $G_d$'s output can be used to produce $G_s$'s outputs.
Finally, we note that \name makes no assumptions about what distribution strategy is used, and can be used with any of them (or with a combination).

\subsection{Example Bugs}\label{s:bg:bugs}
Next, we briefly illustrate the types of bugs that can be introduced when using a distribution strategy to implement a sequential model $G_s$. Later in \S\ref{sec:case-study} we discuss a larger set of bugs and evaluate \name's ability to find them.

\paragraph{Incorrectly scaling auxiliary loss.} In MoE training, auxiliary loss~\cite{gshard,shazeer2017outrageouslylargeneuralnetworks} is used to better balance load among experts by penalizing hot experts. However, when using tensor parallelism, the auxiliary loss needs to be scaled down by the number of TP ranks $T$ (that is, be divided by $T$) to balance out a subsequent reduce-scatter operation that sums up all gradients. We observed a bug at \company where an implementation did not scale-down the auxiliary loss, leading to the distributed implementation producing an auxiliary loss that was $T$-times larger than expected.

\paragraph{Incompatible configurations for model components.} We also observed a bug at \company when switching an MoE model implementation from using TP to shard experts to SP. In this case, the expert weights need to be replicated across SP ranks rather than sharded, but the bug was that MoE weights continued to be sharded rather than replicated when using SP resulting in incorrect output. To illustrate why, consider a sequential model that computes $X\times A$, where $X$ is an input and $A$ are expert weights. SP requires partitioning $X$ into $X_1$ and $X_2$, while sharding partitions $A$ into $A_1, A_2$. The resulting distributed implementation computes $X_1\times A_1$ and $X_2\times A_2$, but these cannot be combined to produce $X\times A$ since the diagonal blocks ($X_1\times A_2$ and $X_2\times A_1$) were never computed. As we explain in \S\ref{sec:case-study} this bug did not change the size of the intermediate data, and thus cannot be caught type checking the model implementation.

\section{Model Refinement}\label{sec:design}
\begin{figure}[t]
    \centering
    \includegraphics[width=\figuresize]{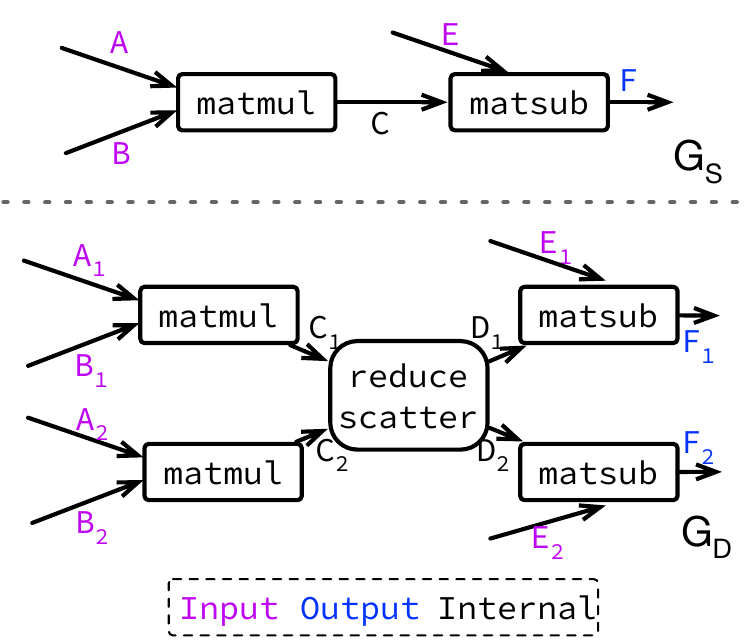}
    \caption{An example of a sequential model $G_s$ and its distributed implementation $G_d$ that is distributed on 2 ranks. $G_s$ produces one output $F$, while $G_d$ produces two outputs $F_1$ and $F_2$. Proving that $G_d$ refines $G_s$ requires finding an expression $\rho$ such that $F=\rho(F_1, F_2)$.}
    \label{fig:graphs}
\end{figure}

In this section, we define the model refinement problem that \name solves. To do so, we first motivate the model refinement problem by giving an overview of how users can use \name (\S\ref{sec:design:overview}) to find bugs in distributed model implementations. Next, in (\S\ref{sec:design:term}), we introduce the notation used in the rest of the paper and formally define the problem. Finally, in \S\ref{s:guarantees} we discuss the guarantees that \name provides when checking model refinement.

\subsection{Overview}\label{sec:design:overview}
Our goal is to check whether a distributed ML model $G_d$ refines a sequential model $G_s$ that is designed to run on a single machine (i.e., uses compute and memory from a single GPU and processor). Note, in most cases using $G_s$ for training or inference is impractical because no single machine might have sufficient resources. However, writing a correct $G_s$ is easier because the model designer does not need to consider communication or coordination across GPUs and ranks.

To use \name, a user provides $G_s$ and $G_d$, which are specified as computation graphs (Figure~\ref{fig:graphs}). Additionally, the user also provides a \emph{clean input relation} (defined below) $R_{i}$ that map $G_s$'s input tensors ($A$, $B$ and $E$ in Figure~\ref{fig:graphs}) to $G_d$'s input tensors ($A_0, A_1, B_0, B_1, E_0, E_1$). From these inputs, \name produces a \emph{clean output relation} $R_o$ that maps $G_s$'s outputs ($F$) to $G_d$'s outputs ($F_0,F_1$). 

From the output relation $R_o$, the user can determine if $G_d$ refines $G_s$ by checking whether $R_o$ is \emph{complete}, that is, does $R_o$ contain mapping for all of $G_s$'s outputs. A complete $R_o$ implies that all of $G_s$'s outputs can be derived from $G_d$'s output without significant computation.
On the other hand, an \emph{incomplete} $R_o$ means that $G_d$ cannot be used to compute at least one of $G_s$'s outputs.

Finally, the user can use a complete $R_o$ to translate outputs from a deployed $G_d$ to $G_s$'s output.

\subsection{Formal Definition and Terminology}\label{sec:design:term}

\begin{table}[ht]
    \begin{center}
        \begin{tabular}{ ll } 
        \toprule
        \textbf{Notation} & \textbf{Explanation} \\
        \midrule
        $\func{I}{(G)}$, $\func{O}{(G)}$ & Set of inputs, outputs of graph G \\
        $\func{I}{(v)}$, $\func{O}{(v)}$ & Set of inputs, outputs of node $v$ \\
        $X\longmapsto Y$ & An expression maps elements in set X to Y \\
        $X\cleanmap Y$ & An expression cleanly maps from set X to Y \\
        $R$ & Relation as a set of tensor-expression pairs \\
        $R_o$ & Clean Relation of outputs \\
        $R_v$ & Clean Relation of outputs of node $v$ \\
        \bottomrule
        \end{tabular}
    \end{center}
    \caption{Notations and explanations}
    \label{table:notations}
\end{table}

The \emph{model refinement problem} requires computing how the sequential model $G_s$'s outputs can be reconstructed from the outputs of the distributed implementation $G_d$:
given model $G_s$ and $G_d$ and a \emph{clean input relation} (defined below) $R_i$ mapping $G_s$'s inputs to $G_d$'s input, solving the model refinement problem requires finding a complete clean output relation $R_o$ that maps all of $G_s$'s output tensors to tensors in $G_d$. If no $R_o$ can be found, model refinement fails, indicating a bug.

In this paper (and in \name), we represent the model $G_s$ and $G_d$ as computation graphs. A computation graph is a directed acyclic graph whose vertices are operators (i.e., computation or communication kernels) and whose edges are tensors. Further, each computation graph $G$ has a set of inputs $I(G)$ and a set of outputs $O(G)$\footnote{For convenience, in our diagrams we represent inputs and outputs as edge that only connect to one vertex.}. We use $T(G)$ to refer to all tensors in a computation graph. Our algorithm requires considering the inputs and outputs of each operator, and we use $I(v)$ (or $O(v)$) to refer to $v$'s input (or output). In \S\ref{s:impl:graph}, we discuss how \name can extract computation graphs from implementations in popular frameworks including PyTorch.

A relation $R$ from computational graph $G$ to $G'$ is a set of tensor-expression pairs: $R = \{(t, \rho) | t\in T(G) \text{ and } t=\rho(T(G')) \}$. An expression is a symbolic description of a computation, and applying expression $\rho$ to an input $x$ evaluates the expression by substituting $x$ for inputs where appropriate. We use the notation $X\longmapsto Y$ to represent an expression mapping elements in the set $X$ to $Y$.

The \emph{input relation} $R_i$ (provided by users) and \emph{output relation} (required as an output) are relations from $G_s$ to $G_d$, $R_i = \{(t, \rho) | t\in I(G_s) \text{ and } t=\rho(I(G_d)) \}$ and $R_o = \{(t, \rho) | t\in O(G_s) \text{ and } t=\rho(O(G_d)) \}$. Each element in $R_i$ (or $R_o$) provides a mapping from $G_s$'s inputs (or outputs) to $G_d$' inputs (or outputs). 
Note, that a relation might provide several mappings for the same tensor $t$, allowing us to model distributed implementations that replicate inputs. 

We define an output relation $R_o$ as \emph{complete} if it contains mappings for all outputs from $G_s$, that is, $\forall o\in O(G_s)\,\exists(o, \rho)\in R_o$.

A \emph{clean expression} $\rho$ is one that consists of two types of operations: \begin{inparaenum}[(i)]
    \item operations including slice, concatenate and transpose that rearrange tensor elements, e.g., by permuting them or masking elements in certain positions;
    \item reduction operations including reduce-sum that perform collective communication and combine tensors distributed across nodes.
\end{inparaenum}
We use the notation $X\cleanmap Y$ to represent a clean expression from set $X$ to $Y$. A \emph{clean output relation} $R_o$ is a relation that contains only clean expressions, i.e., $\forall (t, \rho)\in R_o,\ \rho \text{ is clean}$.

We restrict $R_o$ so that it contains only clean expressions because needing complex computation to reconstruct $G_s$'s outputs from $G_d$ indicates a bug: programmers apply parallelism strategy (\S\ref{s:background}) to create an implementation $G_d$ that is equivalent to the sequential model $G_s$. Combining distributed outputs requires communication and aggregation operations (which clean operations are allowed to perform), but any computation beyond this indicates that $G_d$ is either incomplete or buggy.

\subsection{Assumptions and Guarantees}\label{s:guarantees}
Finally, we list the assumptions made by \name when solving the model refinement problem, and discuss the guarantees that it provides.

\paragraph{Assumptions:}
We make two assumptions about $G_s$ and $G_d$: First, we assume that the 
same set of optimizations (e.g., kernel fusion or using optimized kernels such as FlashAttention~\cite{flashattention}) are applied to both the specification $G_s$ and the implementation $G_d$. Second, we assume that if $G_d$ correctly refines $G_s$, then $G_s$'s outputs can be reconstructed by rearranging or combining $G_d$'s outputs (this assumption is precisely captured by our definition of clean relations). The second assumption is based on our use case: programmers build $G_d$ to implement $G_s$, and thus additional computation to map $G_d$'s outputs to $G_s$ indicate a bug.

\paragraph{Guarantees:} \emph{\name is sound}: that is, if \name says that $G_d$ refines $G_s$, then there exists a clean output relation $R_o$ using which $G_s$'s outputs can be reconstructed from tensors in $G_d$. This is because \name explicitly searches for such an $R_o$, and returns the relation it finds. Thus, \name's output acts as a certificate of soundness.

\emph{\name is not complete}: it can falsely report a bug for a correct $G_d$. This is because \name depends on several assumptions to scale its performance, and it might not find a clean relation if the model or implementation violate these assumptions. Specifically, \name assumes that:
\begin{compactenum}
\item The same optimizations are applied to $G_s$ and $G_d$. This might be violated if a programmer or tool optimizes $G_d$ directly, e.g., by replacing multiple operators by a fused kernel.
\item $G_d$ and $G_s$ perform operations in the same order. As we explain in the next section, this assumption allows us to iteratively verify model refinement, and thus scale to large models.
\item If an operator $v_d\in G_d$ refines operator $v_s\in G_s$ (i.e., $v_s$'s outputs can be computed using $v_d$'s outputs), then $v_d$'s inputs can be mapped to $v_s$'s inputs or outputs. This assumption, which we state more formally in \S\ref{subsec:optimize-exploration}, enables an optimization (\S\ref{subsec:optimize-exploration}) that allows \name to iteratively consider subgraphs of $G_d$ when searching for $R_o$, the clean output relation mapping $G_d$'s outputs to $G_s$'s outputs.
\end{compactenum}

We made these assumptions because we found that they hold for models implemented using several frameworks, including Megatron-LM, vLLM, and \company's internal framework. This is because the parallelism strategies we discussed in \S\ref{s:background} do not require operator fusion or operation reordering. Consequently, we did not find any model implementations that violated these assumptions, and thus did not observe any false bug reports in our evaluation.

However, it is possible that other frameworks or future version of the frameworks we analyzed might violate these requirements, necessitating changes to our approach to allow it to scale. This is similar to other verification tools, that trade-off completeness for performance and scalability.

\section{\name's Approach}\label{sec:approach}
\begin{lstlisting}[language=Python, caption={Algorithm to compute the relations between output tensors of $G_s$ and $G_d$ inductively. We describe the underlined function in \S\ref{sec:step}.},
    float=t,
    belowskip=-0.5 \baselineskip,
    label=alg:inductive, escapechar=`,
     emph={compute_node_out_rel},
    emphstyle=\underbar]
def compute_out_rel(G_s, G_d, R_i):
    sort_vs = topological_sort(G_s) `\label{alg:inductive:sort}`
    R = R_i `\label{alg:inductive:init:r}`
    for v_s in sort_vs:
        R_v = compute_node_out_rel(v_s, G_d, R) `\label{alg:inductive:r_v}`
        if not R_v.contains(O(v_s)): `\label{alg:inductive:error}`
            raise RefinementError("Could not map outputs for operator", v_s)
        R = R `$\cup$` R_v `\label{alg:inductive:upd:r}`
    R_o = {(t_s, expr(T)) | (t_s, expr(T)) `$\in$` R, t_s `$\in$` O(G_s), `$T\subseteq$` O(G_d)} `\label{alg:inductive:ro}`
    return R_o
\end{lstlisting}

\name's approach to computing the clean output relation (Listing~\ref{alg:inductive}) $R_o$ is iterative: it processes each operator $v\in G_s$ (we use $v\in G$ to refer to an operation $v$ in computation graph $G$) in topological order (line~\ref{alg:inductive:sort}) and computes a clean output relation $R_v$ (line~\ref{alg:inductive:r_v}) containing expressions $O(v)\cleanmap T(O(G_d))$ that map $v$'s outputs to $G_d$'s tensors. If $R_v$ is not a complete relation, i.e., it does not contain mappings for all of $v_s$'s outputs, then \name raises an error indicating that $G_d$ does not refine $G_s$ (line~\ref{alg:inductive:error}). The error includes the identity of the operator $v_s$ whose outputs could not be mapped cleanly to $G_d$, enabling bug localization.

On the other hand, if $R_v$ is a complete relation, \name updates $R$, a relation containing clean maps $T(G_s)\cleanmap T(G_d)$ found thus far, by adding the mappings contained in $R_v$  (line~\ref{alg:inductive:upd:r}). Finally, once all operators have been processed, it filters $R$ to produce the clean relation $R_o$ of $O(G_s)\cleanmap O(G_d)$ (line~\ref{alg:inductive:ro}).

The relation $R$ is also provided as an input to the function (Line~\ref{alg:inductive:r_v}) that computes operator $v$'s output relation $R_v$. We describe this function in the next section (\S\ref{sec:step}), and the algorithm needs to map $I(v)$ ($I(v)\subseteq T(G_s)$) into $T(G_d)$. It uses the relation $R$ for this purpose. Processing operators in topological order ensures that this mapping is always feasible: it ensures that the operator being processed either uses tensors in $I(G_d)$ which can be mapped using the user-provided expression $R_i$ that we use as $R$'s initial value (line~\ref{alg:inductive:init:r}) or using the output of previously processed operators whose outputs have been mapped by previous calls to \texttt{compute\_node\_out\_rel}.

The correctness of our approach requires that any tensor in $G_s$ (whether input, intermediate, or output) can be mapped to one or more tensors in $G_d$ by a clean expression: If this requirement is violated, the algorithm would not find a $R_v$ for some operator $v\in G_s$. Our assumption that optimizations (including kernel fusion) applied to $G_d$ must also be applied to $G_s$ ensures that this requirement holds for our inputs.

We illustrate this process using the computation graphs in \Cref{fig:graphs}: initially the algorithm sets $R = R_i$ and processes the \texttt{matmul} operation by finding the relations $R_C=\{(C, \text{ReduceSum}(C_1, C_2)), (C,\texttt{Concat}(D_1, D_2))\}$ (line~\ref{alg:inductive:r_v} returns all mappings for $C$).
that maps the intermediate tensor $C$ to tensors in $G_d$. Next, \name updates $R=R_i\bigcup R_C$ and processes the \texttt{matsub} operation, and finds the relation $R_F=\{(F, \text{Concat}(F_1, F_2, dim=0))\}$. Because $F$ is $G_s$'s only output, \name emits $R_F$ as its output. 

\begin{figure}[t]
    \centering
    \includegraphics[width=\figuresize]{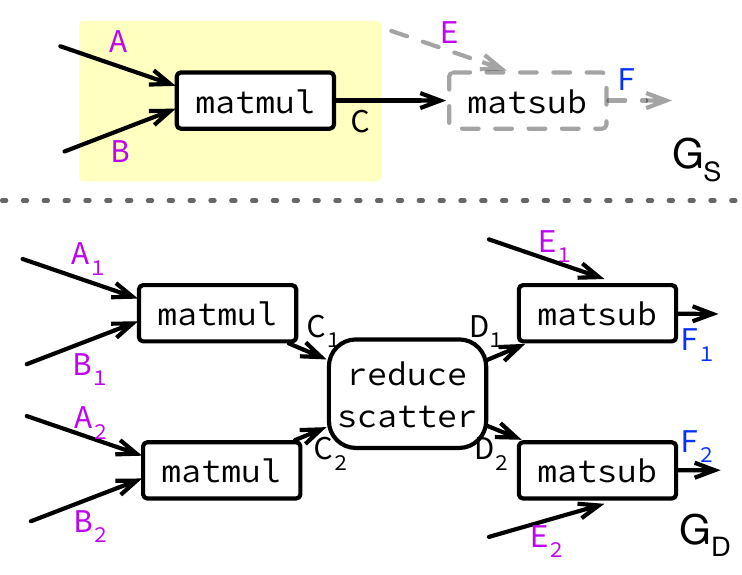}
    \caption{We want to compute the relation between $C$ and all tensors ($A_1, B_1, C_1, D, E_1, F_1, A_2, B_2, C_2, E_2, F_2)$ in $G_d$. The edges are marked with what output tensor is passed to the next node as the input. All tensors here have two dimensions. Assume we have the input relation $R_i$: ($A, \alpha_0:\texttt{concat}(A_1, A_2, \texttt{dim=}1)$), ($B, \beta_0:\texttt{concat}(B_1, B_2, \texttt{dim=}0)$).}
    \label{fig:lemma}
\end{figure}

\subsection{Computing the Output Relation for an Operation}\label{sec:step}
Next, we detail how \name computes the clean output relation $R_v$ for an operator $v\in G_s$ given an input relation $R$ that contains mappings $I(v)\cleanmap T(G_d)$. 

Informally, \texttt{compute\_node\_out\_rel} works as follows: 
\begin{inparaenum}[(i)]
\item it uses the input relation $R$ (provided as input) to produce the expression $\rho_v$ that can be used to compute $v$'s outputs using tensors in $T(G_d)$; 
\item it uses lemmas (\S\ref{subsec:lemma}) to rewrite and find the set of all expressions (\S\ref{sec:proc:rewriting}) $P_v$ that are equivalent to $\rho_v$ (including $\rho_v$); 
\item it uses information from $G_d$ to rewrite and find all equivalent expressions to those appearing in $P_v$, and adds those to $P_v$; and
\item it uses the set of clean expressions in $P_v$ to construct and return the desired output relation $R_v$.
\end{inparaenum}

We illustrate this process by applying it to the \texttt{matmul} operator (Figure~\ref{fig:lemma}) in our running example. Because \texttt{matmul} is the first operator in $G_s$, \texttt{compute\_node\_out\_rel} receives $R=R_i$ as input. The computation proceeds as follows:
\begin{compactenum}[(i)]
\item The algorithm uses input mappings for $A$ and $B$
to produce $\rho_v: \texttt{matmul}(\alpha_0, \beta_0)$  
(where $\alpha_0:\texttt{concat}(A_1, A_2, \texttt{dim}=1)$ and 
$\beta_0:\texttt{concat}(B_1, B_2, \texttt{dim}=0))$). 
\item Next, the algorithm applies the block matrix lemma to rewrite $\rho_v$ and find the equivalent expression $\rho_v^1: \texttt{sum}(\alpha_1, \beta_1)$ ($\alpha_1: \texttt{matmul}(A_1, B_1)$ and $\beta_1: \texttt{matmul}(A_2, B_2))$ ). Another lemma also applies to reduce scatter, but we omit it for clarity.
\item The algorithm uses $G_d$ to find additional rewritings for each $\rho\in P_v$. In this example, it uses the observation that $C_1=\texttt{matmul}(A_1, B_1)$ and $C_2=\texttt{matmul}(A_2, B_2)$ to rewrite $\rho_v^1$ to the equivalent expression $\rho_v^2: \texttt{sum}(C_1, C_2)$. After this step, $P_v$ includes $\{\rho_v, \rho_v^1, \rho_v^2\}$ (along with additional terms from considering reduce-scatter).
\item Finally, the algorithm filters $P_v$ to find clean mappings. In this case $\rho_v^2$ is a clean mapping, and the algorithm returns $R_v = \{(C, \texttt{sum}(C_1, C_2)), (C,\texttt{Concat}(D_1, D_2))\}$, the later of which was computed by considering the reduce-scatter operation.
\end{compactenum}

\subsubsection{Computing $R_v$}\label{subsec:step-algo}
\begin{lstlisting}[language=Python, caption={Algorithm to compute the relations between output tensors of v\_s and all the input/output tensors of a node set G\_d. R is the relations between the input tensors of v and G\_d. The underlined functions are described in \S\ref{sec:proc:rewriting}.},
    float=t,
    belowskip=-0.5 \baselineskip,
    label=alg:lemma,
    escapechar=`,
    emph={rewrite_using_lemma, rewrite_t_to_expr, rewrite_expr_to_t},
    emphstyle=\underbar
    ]
def compute_node_out_rel(v, G_d, R):
    P_v = set()
    base_expr = v(I(v)) `\label{alg:lemma:v1-out}`
    
    # Step 1: replace tensors of I(v_s) with the expressions in R. `\label{alg:lemma:s1:b}`
    P_v = rewrite_t_to_expr(base_expr, R)
    exprs_step1 = P_v `\label{alg:lemma:s1:e}`
    
    # Step 2: rewrite expressions in exprs_step1 based on the given lemmas.`\label{alg:lemma:s2:b}`
    for expr in exprs_step1:
        P_v = P_v `$\cup$` rewrite_using_lemma(expr)
    exprs_step2 = P_v`\label{alg:lemma:s2:e}`
    
    # Step 3: rewrite expressions in exprs_step2 by replacing the sub-expressions with tensors. `\label{alg:lemma:s3:b}`
    T_rel = `$\{t\in T(G_d) | t\text{ is an input of an expression appearing in R}\}$` `\label{alg:lemma:trinit}`
    
    # R_G_d is a relation of all tensors in `$T(G_d)$` that can be computed using tensors in T_rel (including by applying multiple operators). 
    
    R_G_d = expressible_using_tensors(G_d, T_rel) `\label{alg:lemma:rgd}`
    for expr in exprs_step2:
        P_v = P_v `$\cup$` rewrite_expr_to_t(expr, R_G_d) `\label{alg:lemma:s3:e}`
        
    # Step 4: filter the expressions and only keep clean ones
    R_v={(t_s,expr)|t_s`$\in$`O(v_s),expr`$\in$`P_v,`\text{expr is clean}`} `\label{alg:lemma:filter}`
    return R_v
\end{lstlisting}

Listing~\ref{alg:lemma} shows the algorithm used to compute $R_v$ for an operator $v\in G_s$ given the relation $R$ of mappings computed by previous calls to this function. As we discussed above, our approach uses iterative expression rewriting to compute $R_v$. We describe the rewrite functions (which are underlined in the listing) in \S\ref{sec:proc:rewriting}, and present the overall algorithm below:

First, \name computes $v$'s output (line~\ref{alg:lemma:v1-out}) in terms of $G_s$'s tensors, and uses the relation $R$ to express this output in terms of tensors in $T(G_d)$ (lines~\ref{alg:lemma:s1:b}---\ref{alg:lemma:s1:e}), and initializes the set $P_v$ with these expressions. It then applies lemmas to find equivalent expressions, and adds them to $P_v$ (lines~\ref{alg:lemma:s2:b}---\ref{alg:lemma:s2:e}). 

Next, it adds to $P_v$ any expressions produced by rewriting the elements in $P_v$ in terms of tensors in $T(G_d)$ (lines~\ref{alg:lemma:s3:b}---\ref{alg:lemma:s3:e}). Observe that the only tensors in $G_d$ that can appear in $P_v$ must also appear in $R$ because applying lemmas cannot produce an expression accessing additional tensors from $G_d$. 
On line~\ref{alg:lemma:trinit}, we compute this set as \texttt{T\_rel}.
Therefore, to improve efficiency, \name creates a relation \texttt{R\_G\_d} (line~\ref{alg:lemma:rgd}) that map tensors in \texttt{T\_rel} to $T(G_d)$, 
and then use \texttt{R\_G\_d} to rewrite the relations in $P_v$ (line~\ref{alg:lemma:s3:e}).  Finally, it filters $P_v$ to produce its output $R_v$ (line~\ref{alg:lemma:filter}).

\subsection{Rewriting Expressions and Terms}

\subsubsection{Lemma}\label{subsec:lemma}
Expression rewriting is a core part of \name's approach. We depend on rewrite rules, that we refer to as \emph{lemmas} to identify ways to rewrite an expression. As we discuss in \S\ref{sec:impl} \name includes lemmas for common operations in PyTorch's ATen library. Some models rely on optimized kernels or uncommon operators, and we also require users to provide lemmas for this. We evaluate the number of additional lemmas required and the associated effort in \S\ref{sec:lemmas-and-efforts}.

An \name lemma states under what conditions an expression can be rewritten to another. In our exposition, we represent a lemma as 
\(
\rho_m(T_m) \xrightarrow[]{C_m(T_m)} \rho_n(T_n)
\)
This lemma states that the expression $\rho_m(T_m)$ and $\rho_n(T_n)$  are equivalent if $C_m(T_m)$ is true. Consequently, if $C_m(T_m)$ holds, then our algorithm will treat $\rho_n(T_n)$ as a valid rewriting of $\rho_m(T_m)$. It is easy to see if under condition $C_m$ $\rho_m$ can be rewritten to $\rho_n$, there must be some condition $C_n$ under which expression $\rho_n$ can be rewritten as $\rho_m$. Our algorithm assumes that both conversions are available for each lemma, in practice one can generally be derived from the other. 

\subsubsection{Rewriting using \egraphs} \label{sec:proc:rewriting}
Given an operator $v\in G_s$, \name computes the clean output relation $R_v$ by rewriting expressions using lemmas and mapping in the input relation $R$. We use \egraphs (and the \egg~\cite{egg} library) to implement rewriting. Our use of \egg is standard: we represent expressions ($\rho$ above) as \enodes and lemmas as rewrite rules; we run saturation, and then use the resulting \eclasses (containing equivalent relations) in our rewriting functions.

\name uses the following three rewriting functions (listing~\ref{alg:lemma}), all of which return a set of expressions:
\begin{compactitem}
    \item \texttt{rewrite\_on\_lemma}, to find all expressions that can be produced by using lemmas to rewrite the expression $\rho$. When processing $\rho$, this function looks at the \eclass corresponding to $\rho$ and all its subexpressions, and return all expressions equivalent to $\rho$.
    \item \texttt{rewrite\_t\_to\_expr}, which takes as input a relation $R$ and an expression $\rho$, and rewrites variables in $\rho$ using the expressions in $R$. If a tensor $t$ is present in $\rho$ and $(t, \rho_t) \in R$, then this function generates a new expression by replacing every occurrence of $t$ in $\rho$ with $\rho_t$. In our running example, this function is called with the expression $matmul(A, B)$, and a relation containing the tuples $(A, \alpha_0:\texttt{concat}(A_1, A_2, \texttt{dim}=1))$ and $(B, \beta_0:\texttt{concat}(B_1, B_2, \texttt{dim}=1))$, and produces the expression $\texttt{matmul}(\alpha_0, \beta_0)$.
    
    Note this function finds and returns all rewriting, so if two tensors $t$ and $u$ occur in $\rho$, and both $(t, \rho_t) \in R$ and $(u, \rho_u)\in R$ then this function will produce three new expressions from $\rho$: (i) one where $t$ is replaced by $\rho_t$; (ii) one where $v$ is replaced by $\rho_v$; and (iii) finally one where both $t$ and $v$ are replaced. The function will return a set containing all three rewriting and $\rho$.
    
    \item \texttt{rewrite\_expr\_to\_t}, which takes as input a relation $R$ and an expression $\rho$, and replaces sub-expressions of $\rho$ with tensors appearing in $R$ when possible. If $\rho_s$ is a subexpression of $\rho$ and $(s, \rho_s)\in \rho$ then this function creates a new expression by rewriting all occurrences of $\rho_s$ in $\rho$ with $s$. In our running example this function is responsible for rewriting the expression $\texttt{sum}(\alpha_1, \beta_1)$ 
    ($\alpha_1: \texttt{matmul}(A_1, B_1)$ and $\beta_1: \texttt{matmul}(A_2, B_2))$)
    to $\texttt{sum}(C_1, C_2)$ when given a relation $R$ containing the mappings $(C_1, \alpha_1)$ and $(C_2, \alpha_2)$.

    Similar to \texttt{rewrite\_t\_to\_expr}, this function finds and returns all possible rewritings.
\end{compactitem}

\subsection{Optimizations}

\subsubsection{Optimizing Exploration}
\label{subsec:optimize-exploration}

\begin{lstlisting}[language=Python, caption={The optimized version of algorithm that should replace the step 3 (line 18-27) in Listing \ref{alg:lemma}.},
    float=t,
    belowskip=-0.5 \baselineskip,
    firstnumber=14,
    label=alg:optimized, escapechar=`,
     emph={rewrite_using_lemma, rewrite_t_to_expr, rewrite_expr_to_t},
    emphstyle=\underbar]
# Step 3: rewrite expressions in exprs_step2 by replacing the sub-expressions with tensors.
T_rel = `$\{t\in T(G_d) | t\text{ is an input of an expression appearing in R}\}$` `\label{alg:optimized:trinit}`
R_explored = set()
while true:
    # Distinct from R_G_d, R_d only contains tensors in `$T(G_d)$` that can
    # be computed using a single operator, all of whose inputs are in T_rel.
    R_d = `$\{(t, \rho: t\longmapsto \texttt{T\_rel}) | t \text{ is direct children of } \texttt{T\_rel},  \rho\in G_d\}$` `\label{alg:lemma:new-s3-Rdesc}`
    if R_d `$\subseteq$` R_explored:
        break
    R_d = R_d - R_explored `\label{alg:optimized:lemma:b}`
    R_explored = R_explored `$\cup$` R_d `\label{alg:lemma:explored}`
    for expr in exprs_step2:
        P_v = P_v `$\cup$` rewrite_expr_to_t(expr, R_d) `\label{alg:optimized:rewrite-expr}`
    T_rel = T_rel `$\cup$` {t|t is the input of a clean expression in P_v} `\label{alg:optimized:clean-expr}`
\end{lstlisting}

The size of \texttt{R\_G\_d} (line~\ref{alg:lemma:s3:e}) 
has a significant effect on the time taken to process a single operator: \name needs to construct this relation when processing an operator, and it relates to the size of the \egraph used for rewrites. Therefore, we use two observations to further reduce its size.

\paragraph{Observations.} Consider operators $v_s\in G_s$ and $v_d \in G_d$. We observe that in most cases, if $v_s$'s outputs can be cleanly mapped to $v_d$'s outputs, then one of the following two conditions holds for all tensors $t\in I(V_d)$: 
\begin{inparaenum}[(i)]
\item There exists a clean expression that operates on $t$ and maps to a tensor in  $I(v_s)$; or
\item There exists a clean expression that operates on $t$ and maps to a tensor in  $O(v_s)$
\end{inparaenum}

The first condition covers the case where all of $v_d$'s inputs can be mapped to inputs of $v_s$, which indicates that they likely compute related values. The second condition covers the case where a previous operator ($v_p$) already produces output that can mapped to $v_s$'s output, but additional operators are used to further process (e.g., using reduce-scatter it or padding) this output and produce other equivalent outputs. Our goal is to collect all equivalent outputs, necessitating this condition.

Our core observation is that if $v_d$ has inputs that are not related to $v_s$ then $v_d$'s outputs are unlikely to cleanly map to $v_s$'s outputs. Note that these observations hold because of our assumption that the same optimizations were applied to both $G_s$ and $G_d$ (\S\ref{sec:design:overview}). Furthermore, if an input violates this observation, \name will lose completeness (i.e., it might falsely report a bug) but soundness will not be affected (i.e., it will not incorrectly report that model refinement holds).

\paragraph{Using these observations.} Listing~\ref{alg:optimized} shows how we modify line~\ref{alg:lemma:s3:b}---line~\ref{alg:lemma:s3:e} in Listing~\ref{alg:lemma} to reduce the size of the $G_d$ subgraph considered. 

The optimization maintains the set \texttt{T\_rel} of tensors in $G_d$ related to $v$'s inputs or outputs. This set initially contains all tensors $t\in T(G_d)$ that appear in the input relation $R$ (line~\ref{alg:optimized:trinit}). Because we explore the computational graph in topological order, this initial set contains all tensors $t\in G_d$ such that there is an expression that cleanly maps them to $v$'s inputs.

\name then uses an iterative process to find rewritings within the subgraph of $G_d$ that meets the observations: 

\name uses \texttt{R\_d} 
to rewrite expressions found in step 2 (line~\ref{alg:optimized:rewrite-expr}) and adds them to \texttt{P\_v}. \name also adds any tensors $t\in G_d$ that appear in newly added clean expressions in \texttt{P\_v}, and proceeds to the next iteration.

During this process, \name tracks the relations it has considered in each iteration (line~\ref{alg:lemma:explored}), and terminates the process when no additional tensors that meet our observations are identified.

To illustrate how the optimized algorithm works in practice, we revisit the example in \Cref{fig:lemma}.  Initially, \texttt{T\_rel} contains $A_1$, $A_2$, $B_1$, and $B_2$, which are captured in \texttt{R}.

In the first iteration, we consider $C_1$ and $C_2$ form $G_d$, and add $(C_1, \mathtt{matmul}(A_1, B_1))$ and $(C_2, \mathtt{matmul}(A_2, B_2))$ to \texttt{R\_d}. After rewriting, we observe that both $C_1$ and $C_2$ appear as inputs to the clean expression $C = \mathtt{sum}(C_1, C_2)$, and are thus related to $v$'s outputs and are added to \texttt{T\_rel}.

In the next iteration, using the updated \texttt{T\_rel}, which includes $C_1$ and $C_2$, we identify the new tensors $D_1$ and $D_2$ that satisfy our conditions, and we check again whether the expressions can be rewritten using $D_1$ and $D_2$. This second check yields the clean expression $C = \mathtt{concat}(D_1, D_2)$. However, because $E_1, E_2$ are not related to either $A, B$ or $C$, they are not in \texttt{T\_rel}. Therefore, they, and tensors computed using them ($F_1, F_2$) will not be included in \texttt{R\_d}, and thus not be considered in this case.

\subsubsection{Optimizing Term and Expression Rewriting}
\label{subsec:optimize-rewriting}
We also found that naively using EGraphs would produce a large number of rewritten expressions, most of which did not aid in proving model refinement. For example, a lemma of the form $x\rightarrow \texttt{reshape}(\texttt{reshape}(x))$ can be applied to every tensor $t$, and thus produce a large number of rewritten expressions. However, these expressions are generally not useful: \texttt{reshape} is its own inverse, and programmers would not needlessly add extra computation. Thus, we rely on two optimization to reduce the number of unnecessary rewritten expressions that are produced:

\paragraph{Constrained Lemmas.} Some lemmas, e.g., the lemma $X[a:c] \xrightarrow[]{} \texttt{concat}(X[a:b], X[b:c])$ can produce many rewritten expressions because any integer $a<b<c$ is valid. The same is true for the \texttt{reshape} lemma discussed above. However, both lemmas are necessary, these rewrites might be required to prove model refinement, and we cannot remove them. Instead, we add an additional constraint to these lemmas: we require that the target expression or a subexpression (e.g., $\texttt{reshape}(x)$, or both $X[a:b]$ and $X[b:c]$) already appear as \enodes, i.e., they already appear as expression in the computational graph.

\paragraph{Pruning Self-Provable Rewritten Expressions.} Our second optimization does not limit the set of rewritten expressions, but instead limits the set added to a relation $R$. In particular, applying lemmas can produce several equivalent expressions, e.g., $\text{concat}(X[16:32], X[32:48])$ and $X[16:48]$. The equivalence of these expressions is dictated entirely by the rewrite lemmas, and does not depend on any mappings provided by the user or the algorithm. We refer to these equivalent expressions as \emph{self-provable} equivalent expressions. Further, we observe that given one such expression, \texttt{rewrite\_on\_lemma} (\S\ref{sec:proc:rewriting}) can generate all others. Therefore, when maintaining relations (e.g., $P_v$ and $R_v$) we only add the simplest version of each set of self-provable expressions: we pick the expression with the smallest number of nested expression. The reduction in the size of the relation reduces memory requirement and improves performance, without any impact on our tools soundness or completeness.

\section{Implementation and Usage Experience}\label{sec:impl}

We implemented \name in 9000 lines of Python, and the relation inference algorithm in about 7800 lines of Rust code. Of the 7800 lines of Rust, 4100 are used to specify lemmas for PyTorch's ATen library~\cite{core-aten-ir} and to validate the lemmas (e.g., by checking correct shapes and types).
Our implementation includes 92 lemmas, of which 16 were ported from TASO~\cite{taso} and Tensat~\cite{tensat}, while the rest were implemented by us. The lemmas we implemented were based on information about input constraints specified in the PyTorch documentation and on mathematical definitions.

\subsection{Computational Graph Capture}\label{s:impl:graph}
The popularity of deep learning has led to the emergence of numerous model graph capturing tools, that developers use to analyze and optimize models.

In our work, we use \texttt{TorchDynamo}~\cite{torchdynamo}, a model compiler that can capture computation graphs from complex PyTorch model implementations. TorchDynamo uses a torch.fx style graph representations and uses the ATen IR~\cite{core-aten-ir} for common operators (custom operators are represented differently). Most of the models in our evaluation are written using PyTorch.

Although TorchDynamo supports capturing complicated models, we ran into some limitations. The most common issue we encountered was when models used operators whose implementation TorchDynamo did not natively support, in which case the resulting computation graph would have missing operators. Therefore, to ensure that we captured a complete computation graph, model implementer need to adopt the following best practices:
\begin{itemize}
    \item Define \texttt{CustomOp} using the API provided by PyTorch for customized operations.
    \item Replace the original collective communications with the functionalized version\cite{pytorch-functional_collectives} provided by TorchDynamo.
    \item Avoid data-dependent branches.
    \item Disable printing and logging during graph capture.
\end{itemize}
A second issue that we encountered is that the variable names produced by TorchDynamo might differ from the source model because of graph inlining or other operations. To address this, we 
we provide a \texttt{CustomOp} \texttt{log\_tensor} as a helper operation that can record a tensor with a name in the output graphs for any debug purpose.

Our approach is compatible with other frameworks, e.g. TensorFlow~\cite{tensorflow} and JAX~\cite{jax} where XLA~\cite{xla} provides mechanisms to output a model's computation graphs and HLO IR. To demonstrate this, we used \name to check a model implemented using AWS's NeuronX framework that builds on HLO and is compatible with XLA. In this case,
we used XLA to generate the computation graph, and then wrote a utility (in 377 lines of Python code) that translated the output to our intermediate format.

\subsection{Handling Symbolic Scalars}
In computational graphs, tensors do not carry actual data values; instead, they contain only metadata such as shape and data type information. However, certain operators, such as \texttt{select}, can extract individual elements from a tensor, and these extracted scalars can be used to compute the tensor shapes. In TorchDynamo, such scalars are represented as symbolic scalars.

To correctly handle these scalars and apply the rewrites, we sometimes need to reason not only about equality, but also about inequality: for example, some lemmas like the commutativity of $\texttt{concat}(X_1,X_2,dim)[a:b]$ can have different rewriting results depending on the equality (or inequality) of the shapes of $X_1$ and $X_2$, and the values of $a$ and $b$. Consequently, we need to be able to compare these even if they are symbolic. But we cannot compare them if they are symbolic, and we cannot use \egraph for this. Therefore, we encode these scalars using SMT-LIB~\cite{smtlib}. 

Specifically, each scalar in the \egraph is associated with metadata that is either a concrete value or a symbolic identifier. Whenever symbolic comparisons are required, we use SMT-LIB to resolve them using user-specified constraints. This allows \name to verify end-to-end verification over computational graphs that include symbolic values.

\eat{
\subsection{Optimization When Integrating \egg}\label{sec: egg-optimization}
Our algorithm is efficient in exploring relational structures. However, to integrate \egg into our algorithm even more effectively, some additional optimizations were applied. In this section, we present the optimizations employed in our implementation.

\subsubsection*{Bulk Exploration}
Since we implemented our algorithm in the Rust library of \egg, we have to serialize and deserialize the graphs frequently. To avoid time waste in this, we implemented step 1 and 2 together with a single iteration of step 3 of Algorithm~\ref{alg:lemma} in Rust and serialize all necessary data all at once. Then we further augment the exploration part from one step of predecessors (Line \ref{alg:lemma:new-s3-Rdesc} in Algorithm~\ref{alg:optimized} to multiple steps every iteration. \notezhanghan{I may need to rewrite this paragraph after sec4 is done.} \noteding{talk about the cost of multiple iterations and thus we utilize the bulk exploration which helps us to finish exploration with 1 iteration in most cases}
}
\section{Evaluation}

Our evaluation focuses on addressing four questions:

\begin{itemize}
    \item Can \name report bugs efficiently in an informative way when they occur (\cref{sec:case-study})? 
    \item How fast can \name complete an end-to-end verification (\cref{sec:e2e-performance})? And how well does \name scale with respect to the number of parallelism sizes and the layers of models (\cref{sec:scalability})? 
    \item When a user invokes new operators, how much effort is required to complete the operator definition and corresponding lemmas (\cref{sec:lemmas-and-efforts})? 
    \item What lemmas are used when checking model refinement for different models (\cref{sec:lemma-application-analysis})?
\end{itemize}

\subsection{Experiment Setup}
\paragraph{Hardware Setup} We evaluated \name on c6525-25g nodes in CloudLab~\cite{cloudlab}. Each machine has a 16-core AMD EPYC 7302P CPU running at 3GHz, and 128 GB memory. We ran our experiments on Ubuntu-22.04.

\paragraph{Workload Setup} \label{workload-setup}
We evaluate \name using the models shown in 
\Cref{table:framework-model-strategy}. 
Because of limitation in the tool used for capturing computation graphs, most of our evaluation only considered the model's forward pass.
We added instrumentation to an internal model at \company to capture forward, backward and optimizer graphs. The \company internal model is a transformer-based LLM.

We evaluated four commonly used distribution strategies: TP, SP, EP and gradient accumulation. However, as noted previously, our approach does not make assumptions about the distribution strategy and can be applied to others.

We did not evaluate DP and PP, both of which are popular, because of limitations of the graph capturing tool.

For example, in Megatron-LM, DP is optimized with contiguous buffers, which are initialized before the model runs and are not exposed to TorchDynamo. Similarly,
PP relies on intermediate leaf tensors for which it computes gradients, and this is forbidden by TorchDynamo and results in a disconnected graph \cite{dynamo-forbid-requires-grad-in-mid1, dynamo-forbid-requires-grad-in-mid2}.

\begin{table}[t]
    \begin{center}
        \begin{tabular}{lll} 
        \toprule
        \thead{Framework} & \thead{Model} & \thead{Optimization} \\
        \midrule
        \makecell{\company \\ Framework} & \makecell{\company\\Model} & TP, SP, EP \\
        \hline
        Megatron-LM & GPT\footnotemark & TP, SP \\
        \hline
        vLLM & Qwen2 & TP \\
        \hline
        \makecell{Huggingface's \\ transformers} & \makecell{Regression model\\ with MSE\footnotemark} & \makecell{gradient \\ accumulation} \\
        \hline
        Transformers-Neuron & Llama-3 & TP \\
        \bottomrule
        \end{tabular}
    \end{center}
    \caption{A summary of frameworks, models, and optimization strategies.}
    \label{table:framework-model-strategy}
\end{table}
\footnotetext[2]{This is the example GPT training script\cite{megatron-lm-gpt-training-script} in the Megatron-LM repository.}
\footnotetext[3]{This is a test case from Huggingface's transformers repository \cite{hf-regression-model-script}.}

\subsection{Case Study} \label{sec:case-study}
One of our goals in designing \name was to provide users with actionable information when model refinement cannot be proved. We assess this by reproducing 6 real-world bugs and showing how they aid in localizing the problem. Of the bugs we report on, 5 are from \company and 1 is from an open source project. Of these bugs, one in the \company model was found by our tool, ahd  the others had been previously identified.

As we explained in \S\ref{sec:approach}, if \name cannot find $R_o$, it returns the operator $v\in G_s$ where its search terminated. Users can inspect $v$, its input relations, and earlier operators to identify the problem. We illustrate this for each of the bugs below.

\subsubsection{Bugs observed in \company}\hfill\\

\begin{figure}[t]
    \centering
    \includegraphics[width=\figuresize]{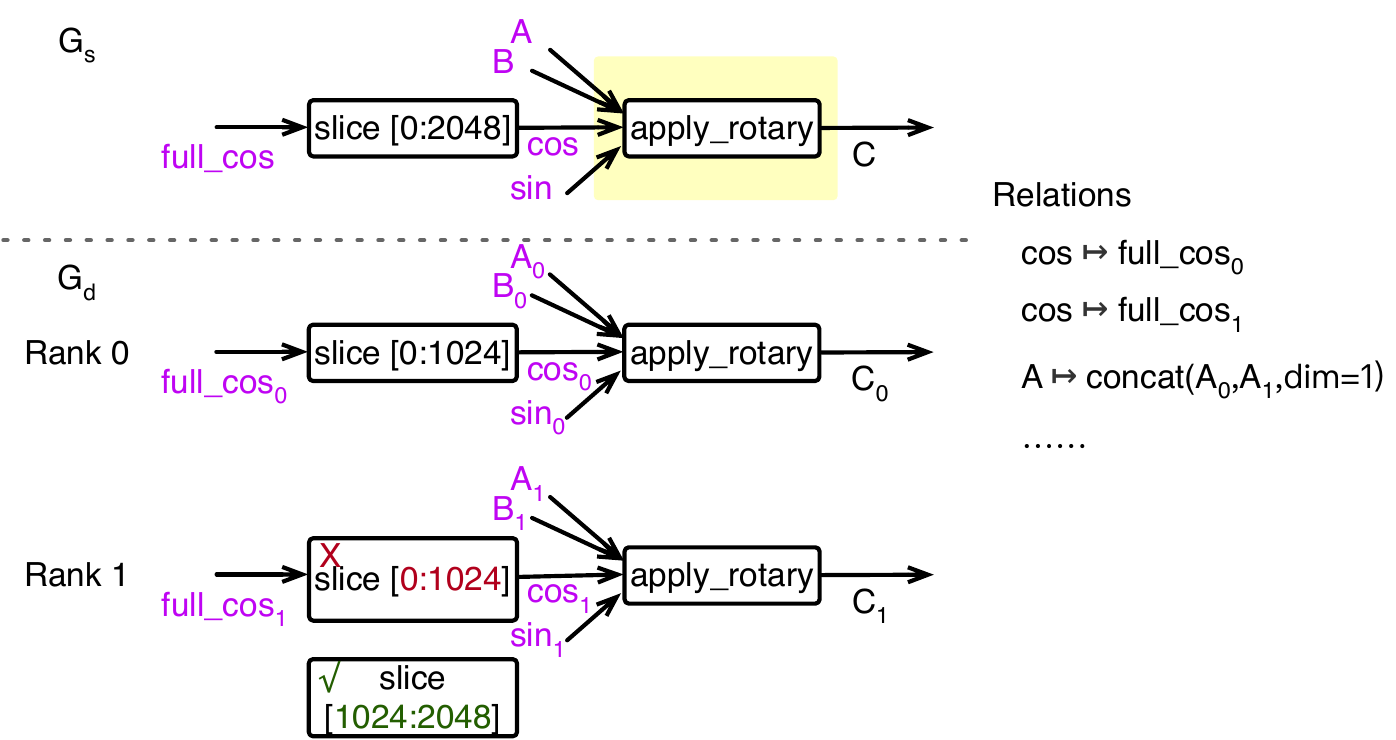}
    \caption{The sub-graphs for the RoPE Bug. Some key relations listed on the right side. }
    \label{fig:rope-bug}
\end{figure}

\paragraph{Bug 1: Incorrect offset in RoPE with SP} When sequence parallelism is enabled, the RoPE embedding~\cite{rope} takes a partition of the original sequence as its input. Consequently, each SP rank should take a different part of the pre-computed $\cos$ and $\sin$ tensors. 
When developing this model, a developer had correctly set the corresponding offset in the code used for the forward pass. However, because the backward pass was implemented using \texttt{torch.autograd.Function} and the developer forgot to also set the corresponding offset in its backward method (\Cref{fig:rope-bug}), resulting in an implementation bug.

\name detect this bug when trying to infer a clean relation for the RoPE operator's output $O(v)$. The user debugs this by checking the input relation $I(v)$ of the operator, which shows that the $\cos$ tensor in $\text{I}{(v)}$ can only be related to the tensor before they are sliced ($\text{full\_cos}_0$ and $\text{full\_cos}_1$ in \Cref{fig:rope-bug}). This is unexpected, because sequence parallelism requires partitioning these tensors and the $\cos$ should also map to $\texttt{concat}(\cos_0,\cos_1,dim=0)$. Tracing back a step further to understand why the mapping between the $\cos$ tensor  and the sliced tensors is missing, the user can see that the slice offsets are incorrect, thus localizing the problem.

\paragraph{Bug 2: Incorrect scaling for auxiliary loss with TP} We mentioned this bug in \S\ref{s:bg:bugs}: During MoE training, auxiliary loss \cite{gshard,shazeer2017outrageouslylargeneuralnetworks} is used to penalize hot experts and improve load balancing. When using TP, the loss should be divided by TP size $T$ to balance out a subsequent reduce-scatter operation that sums up the gradient. Otherwise, the final aggregated gradients can be $T$ times the expected one.

In this case, \name can map outputs for the auxiliary loss update. However, it fails to find a mapping for a subsequent \texttt{matmul} operation that multiples the gradient with another tensor. The bug would require dividing the gradient by TP size to be equivalent, but division is not a clean expression.  In this case, the user works backwards from the \texttt{matmul} operator to identify the missing division.

\paragraph{Bug 3: Mismatched padding and slicing in data processing} 
The \texttt{AllGather} operation that we use at \company requires that the input tensor from senders have the same shape. Thus, when adopting SP, a developer needs to pad tensors to meet this requirement, and subsequently use the \texttt{slice} operator to drop the padding. A bug was introduced when a developer used inconsistent parameters for the padding and slice operators, which resulted in some non-padding elements being dropped and padded elements being retained.

\name detects this bug while inferring clean relation for a subsequent \texttt{baddbmm} operation. In particular, given the operator $v$'s input relation $\text{I}(v)$, \name could not find a clean output relation $\text{O}(v)$. The \name user can inspect both the \texttt{baddbmm} operation and the previous \texttt{slice} operation that produces its inputs to discover that the slice operation had dropped required element. This allows the user to compare parameters for the slice and pad operators, and thus address the problem.

\paragraph{Bug 4: Incompatible configurations for model components}
This bug in \company's model was identified by our tool during this evaluation, and we previous discussed it in \S\ref{s:bg:bugs}. In brief, a developer used SP to parallelize a MoE model, which requires replication the experts' weights. Unfortunately, the developer did not correctly configure some model components, and the expert weights were sharded. This led to a bug: if the sequential model computed $X\times A\times B$, the buggy implementation would compute $X_1\times A_1\times B_1$ and $X_2\times A_2\times B_2$, where $X_1$, $X_2$, $A_1$, $A_2$, $B_1$, $B_2$ are partitions of $X$, $A$ and $B$, respectively. Furthermore, the resulting output still matches the input's hidden dimension size (because the output is still the same size as $X\times A\times B$), and thus the resulting model can be trained. However, the implementation behaves differently from the sequential specification: it never computes the diagonal blocks $X_1\times A_2$ and $X_2\times A_1$, and they do not contribute to the final output.

\name detects this bug when trying to map the first \texttt{matmul}'s output (i.e., $X\times A$) because its output cannot be mapped to any tensor in the implementation. Given this information, the user investigates the operator's input relation and find that input $A$ is incorrectly partitioned.

\paragraph{Bug 5: Missing aggregation for a layernorm weight} 
We observed a bug when deploying a custom distributed optimizer. The developer added a layernorm operation before computing the key tensor in an attention layer, but did not register the layernorm operation's weight with the SP group optimizer. This meant that the layernorm weights were not considered during all-reduce, and thus the gradients computed by this implementation differed from those computed by the sequential model.

In this case \name does not report a bug: it can find a mappings for all of $G_s$'s outputs. However, when the programmer examines the $R_o$ produced by \name, they notice that the relation between the sequential model's gradient and the implementation's gradient differs from the users expectation, indicating a possible bug. The user can then examine the implementation to find the bug.

\subsubsection{Bugs in Open-source Frameworks}
\hfill\\

\paragraph{Bug 6: Wrong scaling in gradient accumulation.} This bug was first reported in 2021~\cite{huggingface-gradient-accumulation-bug-2021} but was misattributed to numeric errors. It was re-reported and finally addressed in 2024~\cite{huggingface-gradient-accumulation-bug-2024}. The bug manifests when gradient accumulation is enabled: gradient accumulation is an approach that splits a batch into multiple min-batches, thus allowing the use larger batch sizes. This approach is similar to the distribution strategies considered above, though the goal is to increase the batch size rather than the number of GPUs. We can also easily obtain a model without gradient accumulation (corresponding to $G_s$) and one with (corresponding to $G_d$), allowing us to use \name.

When using gradient accumulation, the programmer must scale the loss computation for correctness. Otherwise, the computed loss is much larger than would be expected. We evaluated \name's ability to find this bug by creating a simple regression that uses MSE loss.

\name detected this bug because the accumulated loss in $G_d$ cannot cleanly represent the loss in $G_s$ without computation because the loss needs to be scaled by number of accumulation steps in each batch.

\subsection{Verification Time for Different Models} \label{sec:e2e-performance}
Next, we evaluate time taken by \name to compute the output relation and check model refinement. We used \name with the models listed in Table~\ref{table:framework-model-strategy}. The distributed implementation we used had parallelism size set to 2 (i.e., if the model used TP and SP, we would use $2$ TP ranks and $2$ SP ranks), and we checked a single model layer. Checking a single model layer suffices, since all layers have the same operations. Further, we have empirically found that a parallelism size of 2 suffices for finding most bugs. Finally, as we discussed above, because of limitations when capturing computational graphs, for models other than \company's internal model, we only checked the forwards pass.

\begin{figure}[t]
    \centering
    \includegraphics[width=\figuresize]{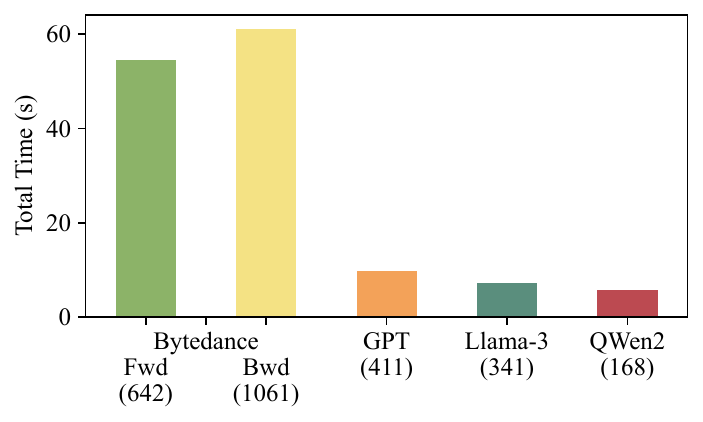}
    \caption{End-to-end verification time across different models. The number in parentheses are the total number of operators in $G_s$ and $G_d$ graphs. The \company "Fwd" and "Bwd" are the forward and backward graphs of \company  proprietary model. }
    \label{fig:end-to-end-time}
\end{figure}

We report times in \Cref{fig:end-to-end-time} for all models other than the HuggingFace's regression model. The Huggingface model was small, and took less than a second. For the remaining models, we observe that verification takes less than 2 minutes for any model (we add the times for X-Fwd and X-Bwd because they represent two passes of the same model), demonstrating that \name can be used while implementing distributed models. Second, we observe that as expected, the verification times are positively correlated with the number of operators used by the model.

\subsection{Scalability} \label{sec:scalability}
Next, we evaluated \name's scalability by measuring verification time as we vary the degree of parallelism and the number of layers (which increases the number of operators). For this evaluation, we used the GPT model and Llama-3. We distributed the GPT model using tensor parallelism (TP), sequence parallelism (SP) and vocabulary parallelism (VP, which is similar to TP) and the Llama-3 model with TP. We used the same degree of parallelism for all types of parallelism.

\begin{figure}[t]
    \centering
    \begin{subfigure}[t]{0.49\linewidth}
        \centering
        \includegraphics[width=\linewidth]{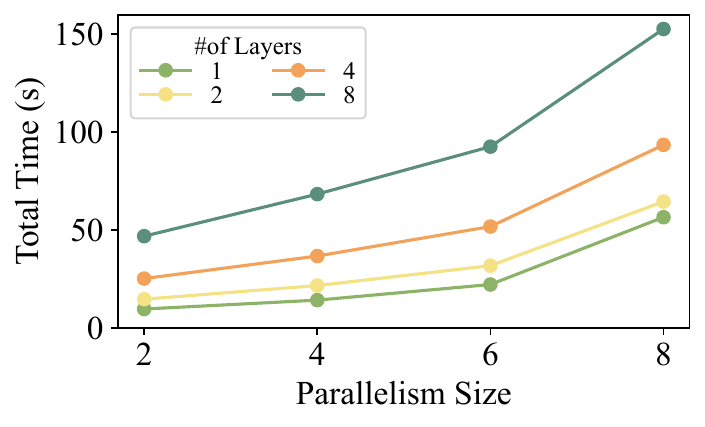}
        \caption{GPT}
    \end{subfigure}
    \begin{subfigure}[t]{0.49\linewidth}
        \centering
        \includegraphics[width=\linewidth]{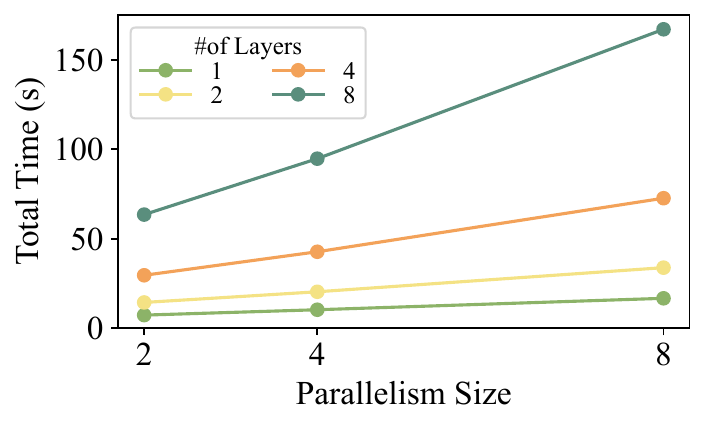}
        \caption{Llama-3}
    \end{subfigure}
    \caption{Scalability on verifying parallelized models. For Llama-3, there is no data for parallelism size as 6, because some component cannot be evenly partitioned by 6.}
    \label{fig:scalability}
\end{figure}

\Cref{fig:scalability} shows the result in terms of verification time as a function of parallelism size and number of layers. We find that using \name remains practical even as the number of operators and parallelism degree increases, showing that it is practical to use our approach in current and emerging deployments. We found that increasing the degree of parallelism has a bigger impact on verification time than increasing the number of layer, but we found that the times remained reasonable up to degree 8. Further, as we observed above, we have also found that for the distribution strategies we used,
increasing the degree of parallelism does not produce additional bugs.

\subsection{Adding Operators and Lemmas} \label{sec:lemmas-and-efforts}
Our implementation (\S\ref{sec:impl}) already includes lemmas for commonly used operators in PyTorch's ATen library. As we show in \S\ref{sec:lemma-application-analysis}, these operators are commonly used. But models also use optimized kernels that fuse operators or are optimized for particular hardware. Furthermore, other IRs, e.g., HLO, provide operators whose semantics might differ from ATen's operators.

\begin{figure}[t]
    \centering
    \begin{subfigure}[t]{0.49\linewidth}
        \centering
        \includegraphics[width=\linewidth]{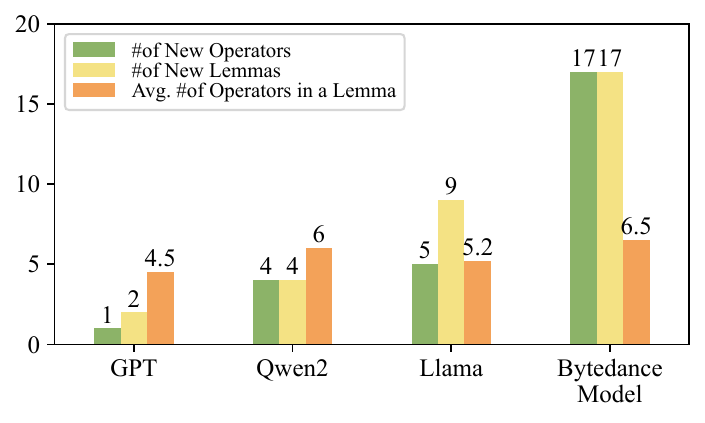}
        \caption{\#of operators and lemmas and average of \#of operators per lemma.}
        \label{fig:efforts:a}
    \end{subfigure}
    \begin{subfigure}[t]{0.49\linewidth}
        \centering
        \includegraphics[width=\linewidth]{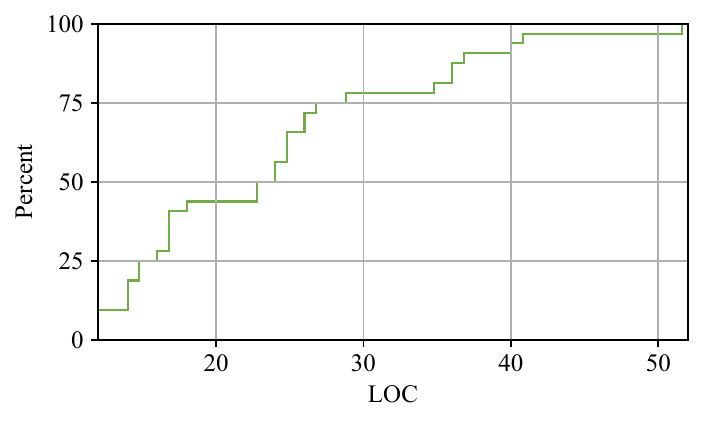}
        \caption{CDF of LOC per lemma.}
        \label{fig:efforts:b}
    \end{subfigure}
    \caption{The efforts to support customized operators. }
    \label{fig:efforts}
\end{figure}

When verifying models that use operators outside the ATen library, \name requires users to provide lemmas that capture the operators semantics. In Figure~\ref{fig:efforts} we quantify the number of new lemmas required to verify the models we evaluated against (Table~\ref{table:framework-model-strategy}) and effort required to create these lemmas.

We quantify effort in two ways: lines of code (shown in the CDF Figure~\ref{fig:efforts:b}), and \emph{lemma complexity}. We measured lemma complexity by counting the number of operators appearing in the lemma. For example, consider the lemma:
\begin{multline*}
\texttt{RMSNorm}(\texttt{concat}(X_1,X_2,dim=0), W)\xrightarrow[]{Cond(X_1, X_2, W)} \\ 
\texttt{concat}(\texttt{RMSNorm}(X_1,W), \texttt{RMSNorm}(X_2,W),dim=0)
\end{multline*}
Two operators appear on the left hand side (\texttt{RMSNorm} and \texttt{concat}) while three appear on the right, and we would assign this lemma a complexity of $5$. For each model, we report the average complexity of all lemmas that were added.

These measurements show that when using \name, users need to add a small number of lemmas, lemmas can be written in a few (< 55) lines of code, and most lemmas are simple. In practice, we found that adding lemmas was not a burden, and thus conclude that the need to add lemmas for optimized operators does not impede \name's usability.

\subsection{Lemma Application Analysis} \label{sec:lemma-application-analysis}

\begin{figure*}[h]
    \centering
    \includegraphics[width=\linewidth]{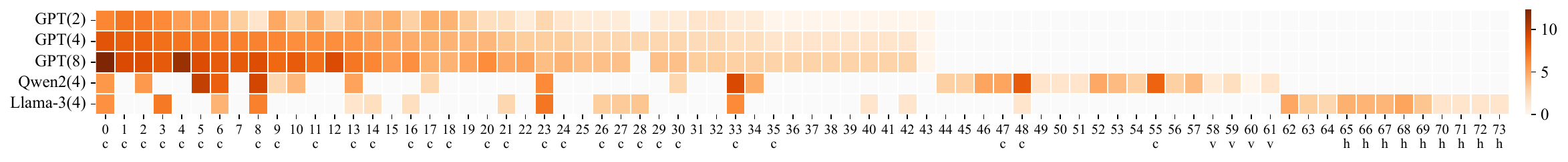}
    \caption{The heatmap shows (in log scale) the number of times each lemma is used for different models, under different parallelism settings. The numbers in parenthesis on the Y-axis represent the degree of parallelism. The x-axis shows lemma IDs, and lemmas marked with \emph{c} concern operators that can appear in clean expressions, those marked with \emph{v} concern operators from vLLM, and those marked with \emph{h} concern HLO operators.}
    \label{fig:lemma_applied_times}
\end{figure*}

Finally, we analyzed the frequency with which different lemmas were used when using \name to check different models. Figure~\ref{fig:lemma_applied_times} shows a heatmap of the number of times each lemma is used. We observe the following from the heatmap:
\begin{itemize}
    \item Lemmas about operators that can appear in clean expressions, including
     \texttt{slice}, and \texttt{concat} are the most commonly used.   
    \item While models that use HLO (Llama-3) require some additional lemmas, they allow us to reuse many of the popular lemmas, including \texttt{slice}, \texttt{reshape} and \texttt{concat} that are developed in the context of ATen.
    \item The different GPT rows show that increasing degree of parallelism increase the number of times lemmas have to be applied. This matches our observation about scalability (\S\ref{sec:scalability}): increasing parallelism has a significant impact on verification time.
\end{itemize}

\section{Related Work}

\paragraph{Verification for ML model transformations.} Prior work has looked at verifying the equivalence between two ML models. Much of this work has been done in the context of superoptimizing ML compilers:
Tensat~\cite{tensat} uses \egraphs to prove equivalence;
TASO~\cite{taso} encodes the equivalence problem in first order logic and uses an SMT solver to check equivalence between models after optimization;
TensorRight~\cite{tensorright} uses rewriting rules to generate bounded proof obligations that can be discharged more efficiently using SMT solvers; and
Mirage~\cite{mirage} and PET~\cite{pet} use a probabilistic approach that evaluates the two models on different inputs to check equivalence. 
The approaches adopted by superoptimizing compilers are generally less scalable than \name's approach, and most compiler only consider rewritings of small subgraphs (e.g., consisting of at most a few tens of nodes). The lack of scalability is because of the different setting that they target: Our scalability builds on the assumption that the output of each operator $v\in G_s$ can be mapped to one or more tensors in $G_d$. But many optimizations, including kernel fusion, that superoptimizing compilers are designed to automate violate this requirement. Consequently, they cannot use \name's iterative approach to scale.

A recent workshop paper~\cite{aerify-euromlsys25} has also proposed Aerify, a system that uses \egraphs to verify semantic equivalence between models. Our work differs in two crucial ways: (a) Aerify's definition of semantic equivalence requires that the both models outputs belong to the same \eclass, i.e., they must be equal. This is a stronger condition than what is required by model refinement: we do not require $G_d$ the produce output that is equal to $G_s$'s output (indeed, this is not the case for many of the implementations we found), but rather that one's output can be mapped to the other. (b) Aerify tries to verify both model optimization and distribution, and similar to verifiers for superoptimizing compilers, our iterative approach cannot be used when reasoning about optimization. Finally, Aerify suggest heuristic model partitioning as a way to scale to larger models, but the use of heuristics often impacts soundness. By contrast, \name is sound.

\paragraph{Fuzz Testing.} Prior work has also proposed using fuzz testing~\cite{LEMON, dl-lib-fuzz, graph-based-fuzz, nnsmith} to evaluate model equivalence. Unlike static analysis based approaches, fuzz testing can scale to large models. However, fuzz testing cannot guarantee soundness. By contrast, verification approaches such as \name can provide soundness guarantees albeit at the cost of scalability.

\paragraph{Optimizing ML Compilers.} Our approach relies on term and expression rewriting. Expression rewriting is central to most ML compilers, including TASO~\cite{taso}, Mirage~\cite{mirage}, TensorRT~\cite{tensorrt}, PET~\cite{pet}, and Mirage~\cite{mirage}. Most of these compilers either develop their own graph substitution algorithms for expression rewriting (TASO, Mirage, etc.), use sketches~\cite{solar2006combinatorial} (TVM, etc.), use expression templates (Taso, TVM, Mirage, etc.), or use a combination. These approaches do not generate \emph{all rewritings} for an expression, and thus cannot be readily used by \name. Tensat~\cite{tensat} is a rewrite of Taso using \egraphs, and thus uses a similar expression rewriting strategy as \name.

\section{Conclusion}
Distributing ML model state and computation across multiple GPUs is a necessity today: model sizes continue to increase, as does the amount of data and compute necessary to train them and use them. We started work on this project because we observed that implementing a distributed model often involved many missteps: bugs would be introduced but go unnoticed in the implementation phased, and would only be noticed during training or later. Formal verification has been used to address similar problems in other domains, e.g., cryptography and networking. But ML differs from these domains in scale: most existing ML models are very large. This led us to design \name, an approach
that uses iterative expression rewriting to  check model refinement and identify bugs introduced when implementing distributed models. The use of iterative expression rewriting allows \name to scale to today's models. 

\clearpage

\bibliographystyle{plainnat}
\bibliography{bibs}

\clearpage


\end{document}